\documentstyle[12pt,epsf]{article} 

\def\mathrm{\rm}

\newcommand {\bea} {\begin{eqnarray}}
\newcommand {\eea} {\end{eqnarray}}
\newcommand {\beq} {\begin{equation}}
\newcommand {\eeq} {\end{equation}}

\begin{document}   
\begin{titlepage}
 \begin{flushright}
  SLAC-PUB-8200\\
  Revised July 1999
%  July/6/98
 \end{flushright}

\begin{center}
 {\Large\bf Direct Measurement of $A_b$ using Charged Kaons
 at the SLD Detector$^*$}
\end{center}

\begin{center}
{\bf The SLD Collaboration$^{**}$}

{\it Stanford Linear Accelerator Center

Stanford University, Stanford, CA 94309 }
\end{center}

\bigskip
\vspace{1.cm} 

\begin{center}
{\bf Contact: Thomas Wright, twright@slac.stanford.edu}
\end{center}

\vspace{1.cm} 

\begin{abstract}
\noindent

%----------------------------------------------------------------------------%
% Abstract                                                                   %
%----------------------------------------------------------------------------%

We report a new measurement of $A_b$ using data obtained by SLD in
1997-98.  This measurement uses a vertex tag technique, where the
selection of a $b$ hemisphere is based on the reconstructed mass of
the bottom hadron decay vertex. The method uses the 3D vertexing
capabilities of SLD's CCD vertex detector and the small and stable SLC
beams to obtain a high $b$-event tagging efficiency and purity of 78\%
and 97\%, respectively.  Charged kaons identified by the CRID detector
provide an efficient quark-antiquark tag, with the analyzing power
calibrated from the data.  We obtain a preliminary result of $A_b =
0.997 \pm 0.044 \pm 0.067$

\end{abstract}

\vfill
 
\begin{center}
{\em Contributed to the International Europhysics Conference on High
Energy Physics, July 15-21 1999, Tampere, Finland; Ref 6-473, and to
the XIXth International Symposium on Lepton Photon Interactions,
August 9-14 1999, Stanford, USA.}
\end{center}

$^*$Work supported by Department of Energy contract
DE--AC03--76SF00515 (SLAC).

\end{titlepage}
\clearpage

\section{Introduction}
 
Measurements of fermion asymmetries at the $Z^0$ resonance probe a
combination of the vector and axial vector couplings of the $Z^0$ to
fermions, $ A_f = 2 v_f a_f / (v_f^2+a_f^2) $.  The parameters $A_{f}$
express the extent of parity violation at the $Zff$ vertex and provide
sensitive tests of the Standard Model.
 
The Born-level differential cross section for the reaction $e^{+}
e^{-} \rightarrow Z^0 \rightarrow f\bar{f}$ is

\begin{equation}
\frac{d \sigma_f}{dz} \propto
(1-A_e P_e) (1+z^2) + 2A_f (A_e - P_e) z \, ,
\end{equation}

where $P_e$ is the longitudinal polarization of the electron beam
($P_e > 0$ for right-handed (R) polarization) and $z = \cos\theta$ is
the direction of the outgoing fermion relative to the incident
electron.  The parameter $A_{f}$ can be isolated by forming the
left-right forward-backward asymmetry ${\tilde{A}}^{f}_{FB}(z) = |P_e|
A_f \, 2z / (1+z^2) \, ,$ although in this analysis we work directly
with the basic cross section.

This note describes the analysis of the data taken during 1997-98 with the
newer VXD3 vertex detector.  Analysis of the 1993-95 data taken with the
original VXD2 vertex detector is described in \cite{abk95}. 

\section{The SLD Detector}
 
The operation of the SLAC Linear Collider with a polarized electron
beam has been described in detail elsewhere \cite{SLC}.  In 1997-98 a
sample of 350k events with average polarization of $ |P_e| = 0.733 \pm
0.008 $ was collected.

Charged particle tracking and momentum analysis are provided by the
Central Drift Chamber \cite{CDC} and the CCD-based vertex detector
\cite{VXD}.  The Liquid Argon Calorimeter (LAC) \cite{LAC} measures
the energy of charged and neutral particles and is also used for
electron identification.  Muon tracking is provided by the Warm Iron
Calorimeter (WIC) \cite{WIC}.  The Cherenkov Ring Imaging Detector
(CRID) \cite{CRID} information (limited to the barrel region) provides
particle identification.  It consists of liquid and gas Cherenkov
radiators illuminating large area UV photon detectors.  The
combination of the two radiators provides good coverage of the
interesting momentum region.

\section{Event Selection} 

Hadronic events are selected based on the visible energy and track
multiplicity in the event. The visible energy is measured using
central drift chamber (CDC) tracks and must exceed 18 GeV.  There must
be at least 7 CDC tracks, 3 with hits in the vertex
detector.  We also require that the thrust axis, measured from
calorimeter clusters, satisfy $|cos{\theta_{thr}}|<0.7$.  This ensures
that the event is contained within the acceptance of the vertex
detector.  All detector elements are also required to be fully
operational. Additionally, we restrict events to 3 jets or less to
make sure that we have well defined hemispheres.  Jets are defined by
the JADE algorithm~\cite{JADE} with $ycut = 0.02$.  A total of 250k
events pass the above hadronic event selection and jet cut.
Background, predominately due to taus, is estimated at $<0.1$\%.

The SLC interaction point (IP) has a size of approximately $(1.5\times
0.5\times 700)$ ${\mu}$m in ($x$,$y$,$z$). The motion of the IP $xy$
position over a short time interval is estimated to be $\sim6$ ${\mu}$m.
Because this motion is smaller than the $xy$ resolution obtained by
fitting tracks to find the primary vertex (PV) in a given event, we
use the average IP position for the $x$ and $y$ coordinates of the
primary vertex.  This average is obtained from tracks with hits in the
vertex detector in ~30 sequential hadronic events.  The $z$ coordinate
of the PV is determined event-by-event.  This results in a
PV uncertainty of $\sim6$ ${\mu}$m transverse and $\sim25$~${\mu}$m
longitudinal to the beam direction.

\subsection{Track Selection}

Reconstruction of the mass of heavy hadrons is initiated by
identifying secondary vertices in each hemisphere. Only tracks that
are well measured are included in the vertex and mass
reconstruction. Tracks are required to have at least 23 CDC hits and
start within a radius of 50 cm of the IP.  The CDC track is also
required to extrapolate to within 1.0 cm of the IP in $xy$ and within
1.5 cm of the PV in $z$.  At least two vertex detector hits are
required, the combined drift chamber + vertex detector fit must
satisfy $\chi^2/{d.o.f.} < 8$, and $|cos\theta|<0.87$. Tracks with an
$xy$ impact parameter $>3.0$ mm or an $xy$ impact parameter error
$>250$ ${\mu}$m with respect to the IP are removed from consideration in
the vertex reconstruction.

\subsection{Vertex Mass Reconstruction}

Vertex identification is done topologically.~\cite{DJNIM}.  This
method searches for space points in 3D where track density functions
overlap. Each track is parameterized by a Gaussian probability density
tube with a width equal to the uncertainty in the measured track
position at the IP.  Points in space where there is a large overlap of
probability density are considered as possible vertex points.  Final
selection of vertices is done by clustering maxima in the overlap
density distribution into vertices for separate hemispheres.  We found
secondary vertices in 86\% of bottom, 45\% of charm, and 2\% of light quark
events.

%Only vertices that are significantly displaced from the PV are
%considered to be possible B or D hadron decay vertices. We require a
%distance between the PV and secondary vertex of at least 1mm.

%Due to the cascade nature of the B decay, tracks from the decay may
%not all originate from the same space point.  Therefore, a process of
%attaching tracks to the secondary vertex has been developed based on
%the transverse and longitudinal distance of closest approach of the
%track to the PV-secondary vertex axis.

The mass of the secondary vertex is calculated using the tracks that
are associated with the vertex.  Each track is assigned the mass of a
charged pion and the invariant mass of the vertex is calculated.  The
reconstructed mass is corrected to account for neutral particles as
follows.  Using kinematic information from the vertex flight path and
the momentum sum of the tracks associated with the secondary vertex,
we add a minimum amount of missing momentum to the invariant mass.
This is done by assuming the true quark momentum is aligned with the
flight direction of the vertex.  The so-called $P_t$-corrected mass is
given by:
$$M_{VTX} = \sqrt{{M^2}_{tk} + {P_t}^2} + |P_t|$$ where $M_{tk}$ is the
mass for the tracks associated with the secondary vertex.  We restrict
the contribution to the invariant mass that the additional transverse
momentum adds to be less than the initial mass of the secondary
vertex. This cut ensures that poorly measured vertices in $uds$ events
do not leak into the sample by adding large $P_t$.

\subsection{Flavor Tag}

We define two tags.  A heavy tag is defined as a hemisphere with an
invariant mass above 2 GeV$/c^2$.  Requiring at least one of these
will ensure high $b$-purity.  The intermediate mass region, between
0.5 and 2 GeV$/c^2$ contains a mixture of $b$ and $c$, with a small
$uds$ background.  The $b$ part of this region is largely $B$'s which
decay so quickly that the track(s) from the $W$ are lost, so only the
$D$ tracks are left.  Because the cascade $K$ that will be used is
among these tracks these vertices are useful, and we define a light
tag to be any hemisphere with a vertex, which doesn't pass the heavy
tag cuts.

\begin{center}
\begin{figure}[htb]
  \epsfxsize12cm
  \hspace{2cm}
  \epsffile{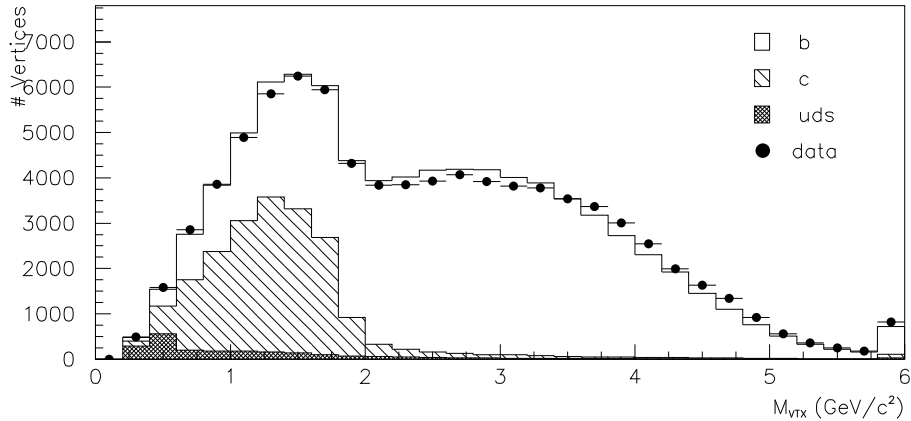}
  \caption{Distribution of $M_{VTX}$ for data (points) and Monte Carlo
           (solid).  The hatched regions represent the light-flavor
           backgrounds.}
  \label{fig_mass}
\end{figure}
\end{center}

These tags are calibrated against the data as described in \cite{rc97}.  The
efficiencies $\eta_b$ (for the light tag) and $\epsilon_b$ (for the heavy
tag), and partial widths $R_b$ and $R_c$ are found by comparing the single-
vs. double-tagged event rates for the two tags.  In addition, the $c$
light tag efficiency $\eta_c$ can be found from the fraction of events with
a light tag in one hemisphere and a heavy tag in the other (mixed tag).  The
light-flavor efficiencies and $\epsilon_c$ are taken from Monte Carlo.

A bottom event is defined to be one with at least one heavy-tagged
hemisphere.  This is found to be
$\sim 78\%$ efficient for bottom events.  With the calibrated
efficiencies and partial widths the bottom purity of these events is
calculated to be $f_c = 96.6\pm0.1\%$.  This is in good agreement with the
Monte Carlo value 96.8\%.  The $c$ background fraction is
2.6\% with $uds$ making up the remainder.

\subsection{Signal Tag}

The determination of the direction of the quark is done using the kaon
charge, $Q_K$. This is the total charge of the CRID-identified kaon
tracks in the vertex.  Because of the cascade nature of the tag the
signals for $b\rightarrow c\rightarrow s$ and $c \rightarrow s$ decays
have the same sign.  This reduces sensitivity to the $c$ background fraction.

\begin{center}
\begin{figure}[htb]
  \epsfxsize12cm
\hspace{2cm}
  \epsffile{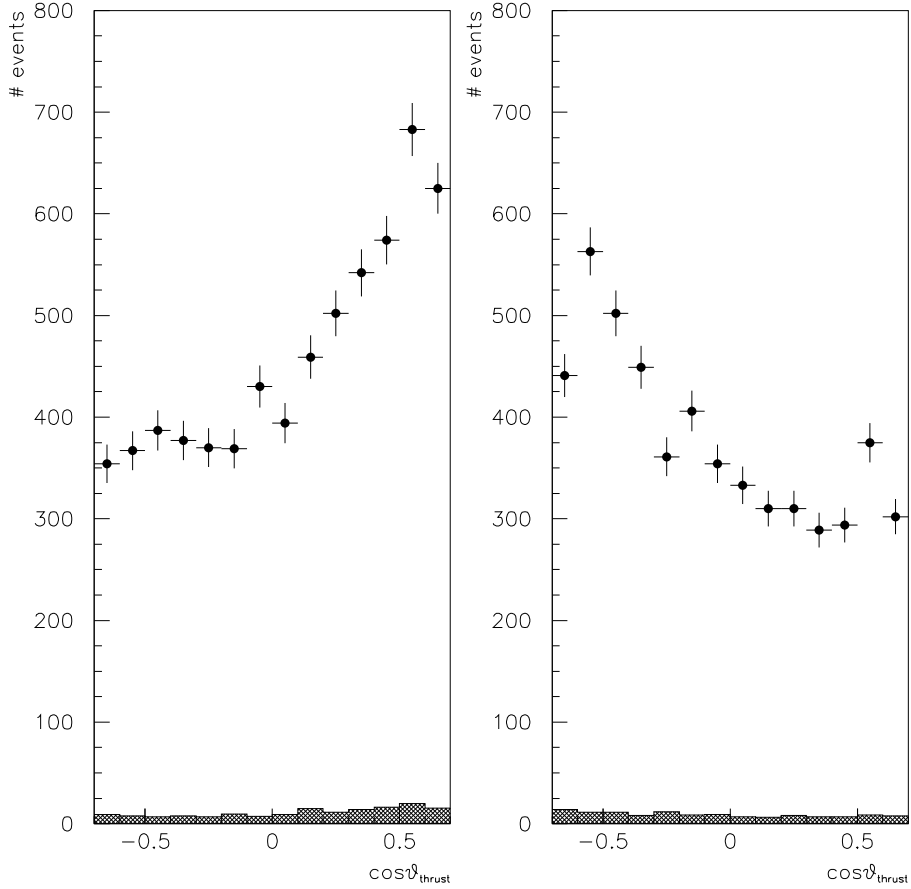}
  \caption{Measured asymmetry for kaon charge.
           Left side is for left-polarized electrons and right 
           side for right-polarized ones.  The hatched region indicates
           the non-$b$ background.}
  \label{fig_acsig}
\end{figure}
\end{center}

We see $\sim25$\% efficiency for the kaon tag for both charm and
bottom events.  A hemisphere with two oppositely-charged $K$
tracks is considered uncharged.

%For
%charm the correct-sign probability is $p_c^{correct}=91$\% and for bottom
%$p_b^{correct}_K=79$\%.  Both tags show a clear asymmetry signal as seen in
%figure \ref{fig_acsig}.

The probability to correctly discrimate between quark/antiquark for
this tag can be calibrated from the data.  The sample used is the
events with both hemispheres $b$-tagged (at least one heavy) and with
nonzero charge in each hemisphere.  The fraction of these events that
are in agreement (opposite charges) can be written as $r_{agree} =
p_{correct}^2 + (1-p_{correct})^2$.  This simply says the hemispheres
must either be both right or both wrong.  After making a correction
for the $c$ contamination we find $p_b^{correct} = 70.7\pm1.4\%$.  The
Monte Carlo gives 72.4\%.  The charm background analyzing power is
taken from Monte Carlo, we find $p_c^{correct} = 84.6\%$.  Because
this procedure calibrates the quark/antiquark flavor at production the
$B$-mixing dilution is automatically included in $p_b^{correct}$.

\section{Results}

   A maximum likelihood fit of all tagged events is used to
determine $A_b$.  As a likelihood function we use the total cross
section:

\begin{eqnarray}
{\cal L} & \propto & (1+z^2)(1-A_e P_e) + 2z(A_e-P_e) \\
& & [ f_b(2p_b^{correct}-1)A_b +  f_c(2p_c^{correct}-1)A_c ] \nonumber
\end{eqnarray}

where $z=-Q\cos\theta_{thr}$, the thrust axis signed by the tagging method
described earlier, is an estimate of the quark direction, $f_{b,c}$ is
the probability for an event to be $b$ or $c$ respectively, and the
factor $(2p^{correct}-1)$ is the effectiveness of the quark/antiquark tag.
The shape of these functions in $z$ is taken from Monte Carlo
with the overall normalizations determined from the data.  The three signs
governing the left-right forward-backward asymmetry -- beam polarization
$P_e$, hemisphere tag charge $Q$, and quark direction $\cos\theta_{thr}$ --
are incorporated automatically into the likelihood function.

%A correction factor $(1-2\chi)$ is applied to all b-quark sources to
%account for asymmetry dilution due to $B^0\bar{B}^0$ mixing, with
%$\chi=.125$ taken from LEP measurements of the average mixing in $Z^0
%\rightarrow b\bar{b}$ events \cite{LEPmix}.

The QCD corrections to the cross-section are well known~\cite{QCDCOR}.
We account for them with a correction term:

\[ 
A^q_{FB|O(\alpha_s)}(\theta) 
= A^q_{FB|O(0)}(\theta) ( 1-\Delta^q_{O(\alpha_s)}(\theta)) 
\]

These QCD corrections have to be adjusted for any bias in the analysis
method against $q\bar q g$ events:

\[ \Delta_{\mbox{QCD}}^{eff} = f  \Delta_{\mbox{QCD}} \]

We estimated the analysis bias factor $f$ for $b$ events from a
generator-level Monte Carlo study:

\[ 
f = \frac{A_{q\bar q}^{gen} - A_{q\bar q+q\bar q g}^{analysis}}
{A_{q\bar q}^{gen} - A_{q\bar q+q\bar q g}^{gen}}
\]

We found $f = 0.89 \pm 0.10$.

From the sample of selected events we measure $A_b = 0.997 \pm 0.044$.

The error is statistical only.

\section{Systematic Errors}

The systematic errors for the 1997-98 SLD result can be found
in Table~\ref{systable}. We give a brief description of the different
sources.

\begin{table}
\caption{Systematic errors for the maximum likelihood analysis}
\label{systable}
\begin{center}
\begin{tabular}{|l|l|}
\hline
Source & $\delta A_b$ \\
\hline
{\bf Tag Composition }&  \\ 
\hline
$f_b$, $f_c$ & 0.001 \\
\hline
{\bf Analyzing Power} & \\
\hline
$p_b^{correct}$ & 0.063\\
Tracking efficiency 3\%& 0.003 \\
MC Statistics & 0.016 \\ 
\hline
{\bf Fit Systematics} & \\ 
\hline
$f,p^{correct}$ shape & 0.011 \\
$\delta P_e$   & 0.010 \\
\hline
{\bf QCD corrections} &   \\ 
\hline
analysis bias & 0.003 \\
$g \rightarrow c\bar c$ & 0.001 \\ 
\hline
{\bf Total} & 0.067  \\ 
\hline
\end{tabular}
\end{center}
\end{table}

The flavor composition error is the statistical error associated with
the efficiency calibration procedure.

The analyzing power error comes mostly from calibration statistics.
Also included here are the tracking efficiency uncertainties and MC
statistics which impact the inter-hemisphere correlation.

The fit systematics include the shape of $f$ and $p$ as functions of
$\cos\theta_{thr}$.  These shapes are taken from the
Monte Carlo and normalized by the calibrated values.  The error is
estimated by fitting with and without these shapes.  Also included in
this category is beam polarization which
is needed in the fit.

The error from QCD corrections comes mainly from the uncertainty in
the correction factor $f$. Gluon splitting is also taken into
account.

\section{Conclusions} We have performed a measurement of $A_b$ using a
method that takes advantage of some of the unique features of the SLC/SLD
experimental program. Our preliminary result based on 250k hadronic $Z^0$
decays is:

\[ A_c = 0.997 \pm 0.044 \pm 0.067 ~~~~~\mbox{\bf Preliminary} \]

This result is consistent with the SM expectation of 0.935 and other
measurements at SLD and LEP.  Because the systematic errors are
dominated by calibration statistics this result is largely
uncorrelated with other measurements.

\section*{Acknowledgements}
We thank the personnel of the SLAC accelerator department and the
technical
staffs of our collaborating institutions for their outstanding efforts
on our behalf.

\vskip .5truecm

\vbox{\footnotesize\renewcommand{\baselinestretch}{1}\noindent
$^*$Work supported by Department of Energy
  contracts:
  DE-FG02-91ER40676 (BU),
  DE-FG03-91ER40618 (UCSB),
  DE-FG03-92ER40689 (UCSC),
  DE-FG03-93ER40788 (CSU),
  DE-FG02-91ER40672 (Colorado),
  DE-FG02-91ER40677 (Illinois),
  DE-AC03-76SF00098 (LBL),
  DE-FG02-92ER40715 (Massachusetts),
  DE-FC02-94ER40818 (MIT),
  DE-FG03-96ER40969 (Oregon),
  DE-AC03-76SF00515 (SLAC),
  DE-FG05-91ER40627 (Tennessee),
  DE-FG02-95ER40896 (Wisconsin),
  DE-FG02-92ER40704 (Yale);
  National Science Foundation grants:
  PHY-91-13428 (UCSC),
  PHY-89-21320 (Columbia),
  PHY-92-04239 (Cincinnati),
  PHY-95-10439 (Rutgers),
  PHY-88-19316 (Vanderbilt),
  PHY-92-03212 (Washington);
  The UK Particle Physics and Astronomy Research Council
  (Brunel, Oxford and RAL);
  The Istituto Nazionale di Fisica Nucleare of Italy
  (Bologna, Ferrara, Frascati, Pisa, Padova, Perugia);
  The Japan-US Cooperative Research Project on High Energy Physics
  (Nagoya, Tohoku);
  The Korea Research Foundation (Soongsil, 1997).}

\bibliographystyle{plain}

%
% author list for inclusion in LaTeX documents
% using \author{} and \address{} commands
%
% Institution number definitions:
%
\section*{$^{**}$ List of Authors}

\begin{center}
\def\iADEL{$^{(1)}$}
\def\iAOMORI{$^{(2)}$}
\def\iBOLO{$^{(3)}$}
\def\iBRI{$^{(4)}$}
\def\iBRUN{$^{(5)}$}
\def\iBU{$^{(6)}$}
\def\iCINC{$^{(7)}$}
\def\iCOLO{$^{(8)}$}
\def\iCOLU{$^{(9)}$}
\def\iCSU{$^{(10)}$}
\def\iFERR{$^{(11)}$}
\def\iFRAS{$^{(12)}$}
\def\iILLI{$^{(13)}$}
\def\iJHU{$^{(14)}$}
\def\iLBL{$^{(15)}$}
\def\iLTU{$^{(16)}$}
\def\iMASS{$^{(17)}$}
\def\iMISSI{$^{(18)}$}
\def\iMIT{$^{(19)}$}
\def\iMOSCOW{$^{(20)}$}
\def\iNAGO{$^{(21)}$}
\def\iOREG{$^{(22)}$}
\def\iOXF{$^{(23)}$}
\def\iPADO{$^{(24)}$}
\def\iPERU{$^{(25)}$}
\def\iPISA{$^{(26)}$}
\def\iRAL{$^{(27)}$}
\def\iRUTG{$^{(28)}$}
\def\iSLAC{$^{(29)}$}
\def\iSOGA{$^{(30)}$}
\def\iSOONG{$^{(31)}$}
\def\iTENN{$^{(32)}$}
\def\iTOHO{$^{(33)}$}
\def\iUCSB{$^{(34)}$}
\def\iUCSC{$^{(35)}$}
\def\iUVIC{$^{(36)}$}
\def\iVAND{$^{(37)}$}
\def\iWASH{$^{(38)}$}
\def\iWISC{$^{(39)}$}
\def\iYALE{$^{(40)}$}

  \baselineskip=.75\baselineskip  
\mbox{Kenji  Abe\unskip,\iNAGO}
\mbox{Koya Abe\unskip,\iTOHO}
\mbox{T. Abe\unskip,\iSLAC}
\mbox{I. Adam\unskip,\iSLAC}
\mbox{T.  Akagi\unskip,\iSLAC}
\mbox{H. Akimoto\unskip,\iSLAC}
\mbox{N.J. Allen\unskip,\iBRUN}
\mbox{W.W. Ash\unskip,\iSLAC}
\mbox{D. Aston\unskip,\iSLAC}
\mbox{K.G. Baird\unskip,\iMASS}
\mbox{C. Baltay\unskip,\iYALE}
\mbox{H.R. Band\unskip,\iWISC}
\mbox{M.B. Barakat\unskip,\iLTU}
\mbox{O. Bardon\unskip,\iMIT}
\mbox{T.L. Barklow\unskip,\iSLAC}
\mbox{G.L. Bashindzhagyan\unskip,\iMOSCOW}
\mbox{J.M. Bauer\unskip,\iMISSI}
\mbox{G. Bellodi\unskip,\iOXF}
\mbox{A.C. Benvenuti\unskip,\iBOLO}
\mbox{G.M. Bilei\unskip,\iPERU}
\mbox{D. Bisello\unskip,\iPADO}
\mbox{G. Blaylock\unskip,\iMASS}
\mbox{J.R. Bogart\unskip,\iSLAC}
\mbox{G.R. Bower\unskip,\iSLAC}
\mbox{J.E. Brau\unskip,\iOREG}
\mbox{M. Breidenbach\unskip,\iSLAC}
\mbox{W.M. Bugg\unskip,\iTENN}
\mbox{D. Burke\unskip,\iSLAC}
\mbox{T.H. Burnett\unskip,\iWASH}
\mbox{P.N. Burrows\unskip,\iOXF}
\mbox{R.M. Byrne\unskip,\iMIT}
\mbox{A. Calcaterra\unskip,\iFRAS}
\mbox{D. Calloway\unskip,\iSLAC}
\mbox{B. Camanzi\unskip,\iFERR}
\mbox{M. Carpinelli\unskip,\iPISA}
\mbox{R. Cassell\unskip,\iSLAC}
\mbox{R. Castaldi\unskip,\iPISA}
\mbox{A. Castro\unskip,\iPADO}
\mbox{M. Cavalli-Sforza\unskip,\iUCSC}
\mbox{A. Chou\unskip,\iSLAC}
\mbox{E. Church\unskip,\iWASH}
\mbox{H.O. Cohn\unskip,\iTENN}
\mbox{J.A. Coller\unskip,\iBU}
\mbox{M.R. Convery\unskip,\iSLAC}
\mbox{V. Cook\unskip,\iWASH}
\mbox{R.F. Cowan\unskip,\iMIT}
\mbox{D.G. Coyne\unskip,\iUCSC}
\mbox{G. Crawford\unskip,\iSLAC}
\mbox{C.J.S. Damerell\unskip,\iRAL}
\mbox{M.N. Danielson\unskip,\iCOLO}
\mbox{M. Daoudi\unskip,\iSLAC}
\mbox{N. de Groot\unskip,\iBRI}
\mbox{R. Dell'Orso\unskip,\iPERU}
\mbox{P.J. Dervan\unskip,\iBRUN}
\mbox{R. de Sangro\unskip,\iFRAS}
\mbox{M. Dima\unskip,\iCSU}
\mbox{D.N. Dong\unskip,\iMIT}
\mbox{M. Doser\unskip,\iSLAC}
\mbox{R. Dubois\unskip,\iSLAC}
\mbox{B.I. Eisenstein\unskip,\iILLI}
\mbox{I.Erofeeva\unskip,\iMOSCOW}
\mbox{V. Eschenburg\unskip,\iMISSI}
\mbox{E. Etzion\unskip,\iWISC}
\mbox{S. Fahey\unskip,\iCOLO}
\mbox{D. Falciai\unskip,\iFRAS}
\mbox{C. Fan\unskip,\iCOLO}
\mbox{J.P. Fernandez\unskip,\iUCSC}
\mbox{M.J. Fero\unskip,\iMIT}
\mbox{K. Flood\unskip,\iMASS}
\mbox{R. Frey\unskip,\iOREG}
\mbox{J. Gifford\unskip,\iUVIC}
\mbox{T. Gillman\unskip,\iRAL}
\mbox{G. Gladding\unskip,\iILLI}
\mbox{S. Gonzalez\unskip,\iMIT}
\mbox{E.R. Goodman\unskip,\iCOLO}
\mbox{E.L. Hart\unskip,\iTENN}
\mbox{J.L. Harton\unskip,\iCSU}
\mbox{K. Hasuko\unskip,\iTOHO}
\mbox{S.J. Hedges\unskip,\iBU}
\mbox{S.S. Hertzbach\unskip,\iMASS}
\mbox{M.D. Hildreth\unskip,\iSLAC}
\mbox{J. Huber\unskip,\iOREG}
\mbox{M.E. Huffer\unskip,\iSLAC}
\mbox{E.W. Hughes\unskip,\iSLAC}
\mbox{X. Huynh\unskip,\iSLAC}
\mbox{H. Hwang\unskip,\iOREG}
\mbox{M. Iwasaki\unskip,\iOREG}
\mbox{D.J. Jackson\unskip,\iRAL}
\mbox{P. Jacques\unskip,\iRUTG}
\mbox{J.A. Jaros\unskip,\iSLAC}
\mbox{Z.Y. Jiang\unskip,\iSLAC}
\mbox{A.S. Johnson\unskip,\iSLAC}
\mbox{J.R. Johnson\unskip,\iWISC}
\mbox{R.A. Johnson\unskip,\iCINC}
\mbox{T. Junk\unskip,\iSLAC}
\mbox{R. Kajikawa\unskip,\iNAGO}
\mbox{M. Kalelkar\unskip,\iRUTG}
\mbox{Y. Kamyshkov\unskip,\iTENN}
\mbox{H.J. Kang\unskip,\iRUTG}
\mbox{I. Karliner\unskip,\iILLI}
\mbox{H. Kawahara\unskip,\iSLAC}
\mbox{Y.D. Kim\unskip,\iSOGA}
\mbox{M.E. King\unskip,\iSLAC}
\mbox{R. King\unskip,\iSLAC}
\mbox{R.R. Kofler\unskip,\iMASS}
\mbox{N.M. Krishna\unskip,\iCOLO}
\mbox{R.S. Kroeger\unskip,\iMISSI}
\mbox{M. Langston\unskip,\iOREG}
\mbox{A. Lath\unskip,\iMIT}
\mbox{D.W.G. Leith\unskip,\iSLAC}
\mbox{V. Lia\unskip,\iMIT}
\mbox{C.Lin\unskip,\iMASS}
\mbox{M.X. Liu\unskip,\iYALE}
\mbox{X. Liu\unskip,\iUCSC}
\mbox{M. Loreti\unskip,\iPADO}
\mbox{A. Lu\unskip,\iUCSB}
\mbox{H.L. Lynch\unskip,\iSLAC}
\mbox{J. Ma\unskip,\iWASH}
\mbox{M. Mahjouri\unskip,\iMIT}
\mbox{G. Mancinelli\unskip,\iRUTG}
\mbox{S. Manly\unskip,\iYALE}
\mbox{G. Mantovani\unskip,\iPERU}
\mbox{T.W. Markiewicz\unskip,\iSLAC}
\mbox{T. Maruyama\unskip,\iSLAC}
\mbox{H. Masuda\unskip,\iSLAC}
\mbox{E. Mazzucato\unskip,\iFERR}
\mbox{A.K. McKemey\unskip,\iBRUN}
\mbox{B.T. Meadows\unskip,\iCINC}
\mbox{G. Menegatti\unskip,\iFERR}
\mbox{R. Messner\unskip,\iSLAC}
\mbox{P.M. Mockett\unskip,\iWASH}
\mbox{K.C. Moffeit\unskip,\iSLAC}
\mbox{T.B. Moore\unskip,\iYALE}
\mbox{M.Morii\unskip,\iSLAC}
\mbox{D. Muller\unskip,\iSLAC}
\mbox{V. Murzin\unskip,\iMOSCOW}
\mbox{T. Nagamine\unskip,\iTOHO}
\mbox{S. Narita\unskip,\iTOHO}
\mbox{U. Nauenberg\unskip,\iCOLO}
\mbox{H. Neal\unskip,\iSLAC}
\mbox{M. Nussbaum\unskip,\iCINC}
\mbox{N. Oishi\unskip,\iNAGO}
\mbox{D. Onoprienko\unskip,\iTENN}
\mbox{L.S. Osborne\unskip,\iMIT}
\mbox{R.S. Panvini\unskip,\iVAND}
\mbox{C.H. Park\unskip,\iSOONG}
\mbox{T.J. Pavel\unskip,\iSLAC}
\mbox{I. Peruzzi\unskip,\iFRAS}
\mbox{M. Piccolo\unskip,\iFRAS}
\mbox{L. Piemontese\unskip,\iFERR}
\mbox{K.T. Pitts\unskip,\iOREG}
\mbox{R.J. Plano\unskip,\iRUTG}
\mbox{R. Prepost\unskip,\iWISC}
\mbox{C.Y. Prescott\unskip,\iSLAC}
\mbox{G.D. Punkar\unskip,\iSLAC}
\mbox{J. Quigley\unskip,\iMIT}
\mbox{B.N. Ratcliff\unskip,\iSLAC}
\mbox{T.W. Reeves\unskip,\iVAND}
\mbox{J. Reidy\unskip,\iMISSI}
\mbox{P.L. Reinertsen\unskip,\iUCSC}
\mbox{P.E. Rensing\unskip,\iSLAC}
\mbox{L.S. Rochester\unskip,\iSLAC}
\mbox{P.C. Rowson\unskip,\iCOLU}
\mbox{J.J. Russell\unskip,\iSLAC}
\mbox{O.H. Saxton\unskip,\iSLAC}
\mbox{T. Schalk\unskip,\iUCSC}
\mbox{R.H. Schindler\unskip,\iSLAC}
\mbox{B.A. Schumm\unskip,\iUCSC}
\mbox{J. Schwiening\unskip,\iSLAC}
\mbox{S. Sen\unskip,\iYALE}
\mbox{V.V. Serbo\unskip,\iSLAC}
\mbox{M.H. Shaevitz\unskip,\iCOLU}
\mbox{J.T. Shank\unskip,\iBU}
\mbox{G. Shapiro\unskip,\iLBL}
\mbox{D.J. Sherden\unskip,\iSLAC}
\mbox{K.D. Shmakov\unskip,\iTENN}
\mbox{C. Simopoulos\unskip,\iSLAC}
\mbox{N.B. Sinev\unskip,\iOREG}
\mbox{S.R. Smith\unskip,\iSLAC}
\mbox{M.B. Smy\unskip,\iCSU}
\mbox{J.A. Snyder\unskip,\iYALE}
\mbox{H. Staengle\unskip,\iCSU}
\mbox{A. Stahl\unskip,\iSLAC}
\mbox{P. Stamer\unskip,\iRUTG}
\mbox{H. Steiner\unskip,\iLBL}
\mbox{R. Steiner\unskip,\iADEL}
\mbox{M.G. Strauss\unskip,\iMASS}
\mbox{D. Su\unskip,\iSLAC}
\mbox{F. Suekane\unskip,\iTOHO}
\mbox{A. Sugiyama\unskip,\iNAGO}
\mbox{S. Suzuki\unskip,\iNAGO}
\mbox{M. Swartz\unskip,\iJHU}
\mbox{A. Szumilo\unskip,\iWASH}
\mbox{T. Takahashi\unskip,\iSLAC}
\mbox{F.E. Taylor\unskip,\iMIT}
\mbox{J. Thom\unskip,\iSLAC}
\mbox{E. Torrence\unskip,\iMIT}
\mbox{N.K. Toumbas\unskip,\iSLAC}
\mbox{T. Usher\unskip,\iSLAC}
\mbox{C. Vannini\unskip,\iPISA}
\mbox{J. Va'vra\unskip,\iSLAC}
\mbox{E. Vella\unskip,\iSLAC}
\mbox{J.P. Venuti\unskip,\iVAND}
\mbox{R. Verdier\unskip,\iMIT}
\mbox{P.G. Verdini\unskip,\iPISA}
\mbox{D.L. Wagner\unskip,\iCOLO}
\mbox{S.R. Wagner\unskip,\iSLAC}
\mbox{A.P. Waite\unskip,\iSLAC}
\mbox{S. Walston\unskip,\iOREG}
\mbox{S.J. Watts\unskip,\iBRUN}
\mbox{A.W. Weidemann\unskip,\iTENN}
\mbox{E. R. Weiss\unskip,\iWASH}
\mbox{J.S. Whitaker\unskip,\iBU}
\mbox{S.L. White\unskip,\iTENN}
\mbox{F.J. Wickens\unskip,\iRAL}
\mbox{B. Williams\unskip,\iCOLO}
\mbox{D.C. Williams\unskip,\iMIT}
\mbox{S.H. Williams\unskip,\iSLAC}
\mbox{S. Willocq\unskip,\iMASS}
\mbox{R.J. Wilson\unskip,\iCSU}
\mbox{W.J. Wisniewski\unskip,\iSLAC}
\mbox{J. L. Wittlin\unskip,\iMASS}
\mbox{M. Woods\unskip,\iSLAC}
\mbox{G.B. Word\unskip,\iVAND}
\mbox{T.R. Wright\unskip,\iWISC}
\mbox{J. Wyss\unskip,\iPADO}
\mbox{R.K. Yamamoto\unskip,\iMIT}
\mbox{J.M. Yamartino\unskip,\iMIT}
\mbox{X. Yang\unskip,\iOREG}
\mbox{J. Yashima\unskip,\iTOHO}
\mbox{S.J. Yellin\unskip,\iUCSB}
\mbox{C.C. Young\unskip,\iSLAC}
\mbox{H. Yuta\unskip,\iAOMORI}
\mbox{G. Zapalac\unskip,\iWISC}
\mbox{R.W. Zdarko\unskip,\iSLAC}
\mbox{J. Zhou\unskip.\iOREG}

\it
  \vskip \baselineskip                   % \bigskip did not work
  \centerline{(The SLD Collaboration)}   % include collaboration name
  \vskip \baselineskip        
  \baselineskip=.75\baselineskip   % shrink the interline spacing
\iADEL
  Adelphi University, Garden City, New York 11530, \break
\iAOMORI
  Aomori University, Aomori , 030 Japan, \break
\iBOLO
  INFN Sezione di Bologna, I-40126, Bologna, Italy, \break
\iBRI
  University of Bristol, Bristol, U.K., \break
\iBRUN
  Brunel University, Uxbridge, Middlesex, UB8 3PH United Kingdom, \break
\iBU
  Boston University, Boston, Massachusetts 02215, \break
\iCINC
  University of Cincinnati, Cincinnati, Ohio 45221, \break
\iCOLO
  University of Colorado, Boulder, Colorado 80309, \break
\iCOLU
  Columbia University, New York, New York 10533, \break
\iCSU
  Colorado State University, Ft. Collins, Colorado 80523, \break
\iFERR
  INFN Sezione di Ferrara and Universita di Ferrara, I-44100 Ferrara, Italy, \break
\iFRAS
  INFN Lab. Nazionali di Frascati, I-00044 Frascati, Italy, \break
\iILLI
  University of Illinois, Urbana, Illinois 61801, \break
\iJHU
  Johns Hopkins University,  Baltimore, Maryland 21218-2686, \break
\iLBL
  Lawrence Berkeley Laboratory, University of California, Berkeley, California 94720, \break
\iLTU
  Louisiana Technical University, Ruston,Louisiana 71272, \break
\iMASS
  University of Massachusetts, Amherst, Massachusetts 01003, \break
\iMISSI
  University of Mississippi, University, Mississippi 38677, \break
\iMIT
  Massachusetts Institute of Technology, Cambridge, Massachusetts 02139, \break
\iMOSCOW
  Institute of Nuclear Physics, Moscow State University, 119899, Moscow Russia, \break
\iNAGO
  Nagoya University, Chikusa-ku, Nagoya, 464 Japan, \break
\iOREG
  University of Oregon, Eugene, Oregon 97403, \break
\iOXF
  Oxford University, Oxford, OX1 3RH, United Kingdom, \break
\iPADO
  INFN Sezione di Padova and Universita di Padova I-35100, Padova, Italy, \break
\iPERU
  INFN Sezione di Perugia and Universita di Perugia, I-06100 Perugia, Italy, \break
\iPISA
  INFN Sezione di Pisa and Universita di Pisa, I-56010 Pisa, Italy, \break
\iRAL
  Rutherford Appleton Laboratory, Chilton, Didcot, Oxon OX11 0QX United Kingdom, \break
\iRUTG
  Rutgers University, Piscataway, New Jersey 08855, \break
\iSLAC
  Stanford Linear Accelerator Center, Stanford University, Stanford, California 94309, \break
\iSOGA
  Sogang University, Seoul, Korea, \break
\iSOONG
  Soongsil University, Seoul, Korea 156-743, \break
\iTENN
  University of Tennessee, Knoxville, Tennessee 37996, \break
\iTOHO
  Tohoku University, Sendai 980, Japan, \break
\iUCSB
  University of California at Santa Barbara, Santa Barbara, California 93106, \break
\iUCSC
  University of California at Santa Cruz, Santa Cruz, California 95064, \break
\iUVIC
  University of Victoria, Victoria, British Columbia, Canada V8W 3P6, \break
\iVAND
  Vanderbilt University, Nashville,Tennessee 37235, \break
\iWASH
  University of Washington, Seattle, Washington 98105, \break
\iWISC
  University of Wisconsin, Madison,Wisconsin 53706, \break
\iYALE
  Yale University, New Haven, Connecticut 06511. \break

\rm
%
%  }   % end of address list

\end{center}

\end{document}